\documentclass[a4paper]{jpconf}
\usepackage{graphicx}
\usepackage{wrapfig}
\usepackage{subfig}
\usepackage{sidecap}
\usepackage{amsmath}
\usepackage{amssymb}
\usepackage{lineno}
\graphicspath{{figures/}}
\begin{document}
\title{Production of $\pi^{0}$, $K^{\pm}$ and $\eta$ mesons in Pb-Pb and pp collisions at $\sqrt{s_{\rm NN}}=$2.76~TeV measured with the ALICE detector at the LHC}

\author{Astrid Morreale on behalf of the ALICE collaboration}

\address{Laboratoire de Physique Subatomique et des Technologies Associ\'{e}es SUBATECH.\\
4 Rue Alfred Kastler, 44300 Nantes, France}

\ead{astrid.morreale@cern.ch}
% \linenumbers
\begin{abstract}
One of the key signatures of the Quark-Gluon Plasma (QGP) is the modification of hadron transverse momentum differential cross-sections in heavy-ion collisions (HIC) as compared to proton-proton (pp) collisions.  Suppression of hadron production at high transverse momenta ($p_{\rm T}$)~in HIC has been explained by the energy loss of the partons produced in the hard scattering processes which traverse the deconfined quantum chromodynamic (QCD) matter. The dependence of the observed suppression on the $p_{\rm T}$~ of the measured hadron towards higher $p_{\rm T}$~  is an important input for the theoretical understanding of jet quenching effects in the QGP and the nature of the energy loss, while suppression  towards lower $p_{\rm T}$ gives information about collective behaviour.
The ALICE experiment at the Large Hadron Collider (LHC) performs a variety of measurements including the spectra of neutral mesons and kaons at mid-rapidity in a wide $p_{\rm T}$~ range in pp, p-Pb and Pb-Pb collisions.
%Neutral mesons ($\pi^{0}$, $\eta$, $\omega$) are reconstructed via complementary methods, using the ALICE electromagnetic calorimeters, PHOS and EMCal, and by the central tracking system, identifying photons converted into $e^{+}e^{-}$ pairs in the material of the inner barrel detectors: the Time projection Chamber(TPC) and the Inner Tracking System  (ITS). Kaon particle identification is performed using  the TPC, the ITS as well as the Time of Flight system (TOF). 
An overview of ALICE results in HIC and pp collisions  at $\sqrt{s_{NN}}=$2.76~TeV of neutral pions, kaons and eta mesons as a function of transverse momentum ($p_{\rm T}$) and centrality will be presented. Ratios $\eta$/$\pi^{0}$, $K^{\pm}$/$\pi^{\pm}$ as well as comparisons to model calculations will also be discussed.
\end{abstract}
\vspace{-1.0cm}
\section{Introduction}
Light mesons offer an opportunity to study the QGP and quantify its properties. At low $p_{\rm T}$, mesons give insight about bulk properties and collective effects. Particles with different masses are expected to be affected differently by the collective motion of the medium; one would thus expect a mass ordering of the nuclear modification factor $R_{\rm AA}$ at low $p_{\rm T}$ due to radial flow.
At high $p_{\rm T}$, hadrons result from the hadronization of partons created in initial hard scattering.
The suppression of hadron production at high $p_{\rm T}$ in heavy-ion collisions is interpreted as energy loss of the scattered parton in the QGP. Gluons are expected to suffer a larger energy loss in the medium than quarks (Casimir factor). 
This effect may give rise to differences in suppression patterns between $\pi^{0}$~ and $\eta$ which can be quantified with $R_{\rm AA}$ measurements.
While one would expect a larger energy loss of high $p_{\rm T}$ partons at higher $\sqrt{s}$, a smaller suppression may be observed at high $p_{\rm T}$
due to a less steeply falling spectrum for a given magnitude of partonic energy loss. A smaller suppression at higher $p_{\rm T}$ may additionally indicate a larger fractional quark to gluon partonic contribution to the production of mesons.
\vspace{-0.3cm}
\section{ALICE's neutral meson reconstruction and charged particle detection}
The ALICE detector~\cite{ALICE} measures photons with three fully complementary methods: photon conversion method (PCM) and electromagnetic shower calorimetry using two detectors: PHOS~\cite{PHOS} which covers a pseudorapidity range of 0.26 units and a coverage in azimuth of $60^\circ$  and the EMCal~\cite{Abelev:2014ffa} with a pseudorapidity and azimuth coverage of  1.4 units and $100^\circ$ respectively. The electromagnetic calorimeters are based on energy measurements via total absorption of particles.

The PCM method measures photons and meson yields by reconstructing $e^{+}e^{-}$ pairs proceeding from photon conversions in the material of ALICE inner detectors. PCM measurements use the ALICE Inner Tracking System (ITS~\cite{ALICE_ITS}) and the Time Projection Chamber (TPC~\cite{ALICE_TPC}), measurements using this method have full azimuthal acceptance and 1.8 units of pseudo rapidity coverage.

\setlength{\intextsep}{-14pt}%
\setlength{\columnsep}{8pt}%
\begin{wrapfigure}[11]{l}{0.42\textwidth}
\vspace*{0.1cm}
\begin{center}
\includegraphics[scale=0.2]{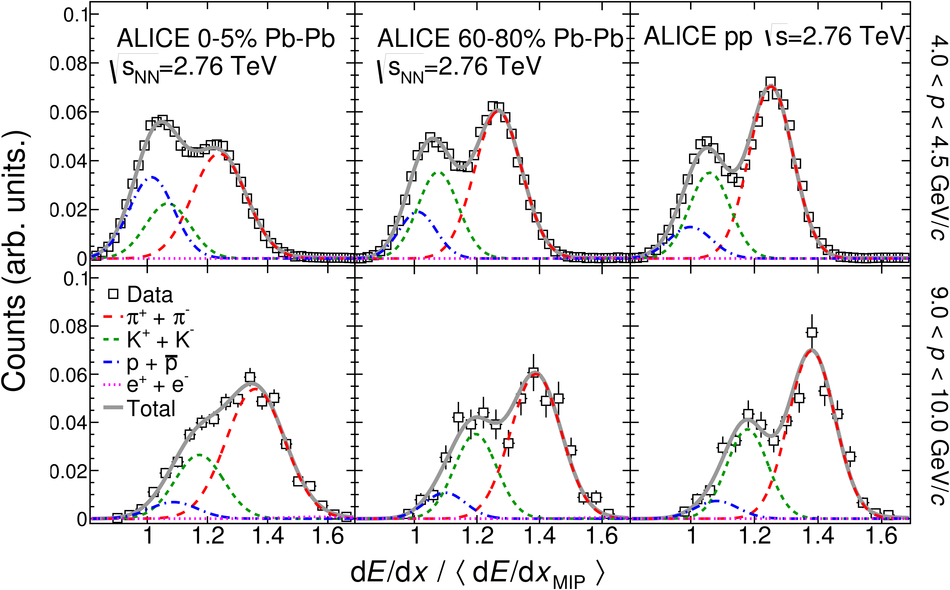}
\caption{\footnotesize d$E$/$d$x distributions of charged particle signals in Pb-Pb (two leftmost columns) and pp (right column). Signals are fitted with the sum of four Gaussian functions (solid line). Top row corresponds to particles with 4.0$~<p<~$4.5 GeV/$c$ while bottom row corresponds to 9.0$~<p<~$10 GeV/$c$.}\label{fig:performanceEnergyloss}
\end{center}
\vspace*{0.0cm}
\end{wrapfigure}

Charged particle identification is  done via the specific energy loss (d$E$/d$x$) of particles traversing the TPC gas.
The candidate signals are fitted with the sum of four Gaussian functions (Fig.~\ref{fig:performanceEnergyloss}) where the remaining contamination from electrons remains below 1$\%$. Particle identification is further improved at p$_{\rm T}<$~4~ ${\rm GeV}/c$ ($K^{\pm}$) with the use of the Time of Flight (TOF) and the High Momentum Particle Identification Detector (HMPID).  Charged particles are detected up to p$_{T}=$~20 GeV/$c$ and at mid-rapidity $\mid \eta \mid \lesssim$~0.8.          

\setlength{\intextsep}{-14pt}
\setlength{\columnsep}{8pt} 
 
\begin{wrapfigure}[32]{l}{0.42\textwidth}
\vspace{2.0cm}
\begin{center}
 
\includegraphics[scale=0.40]{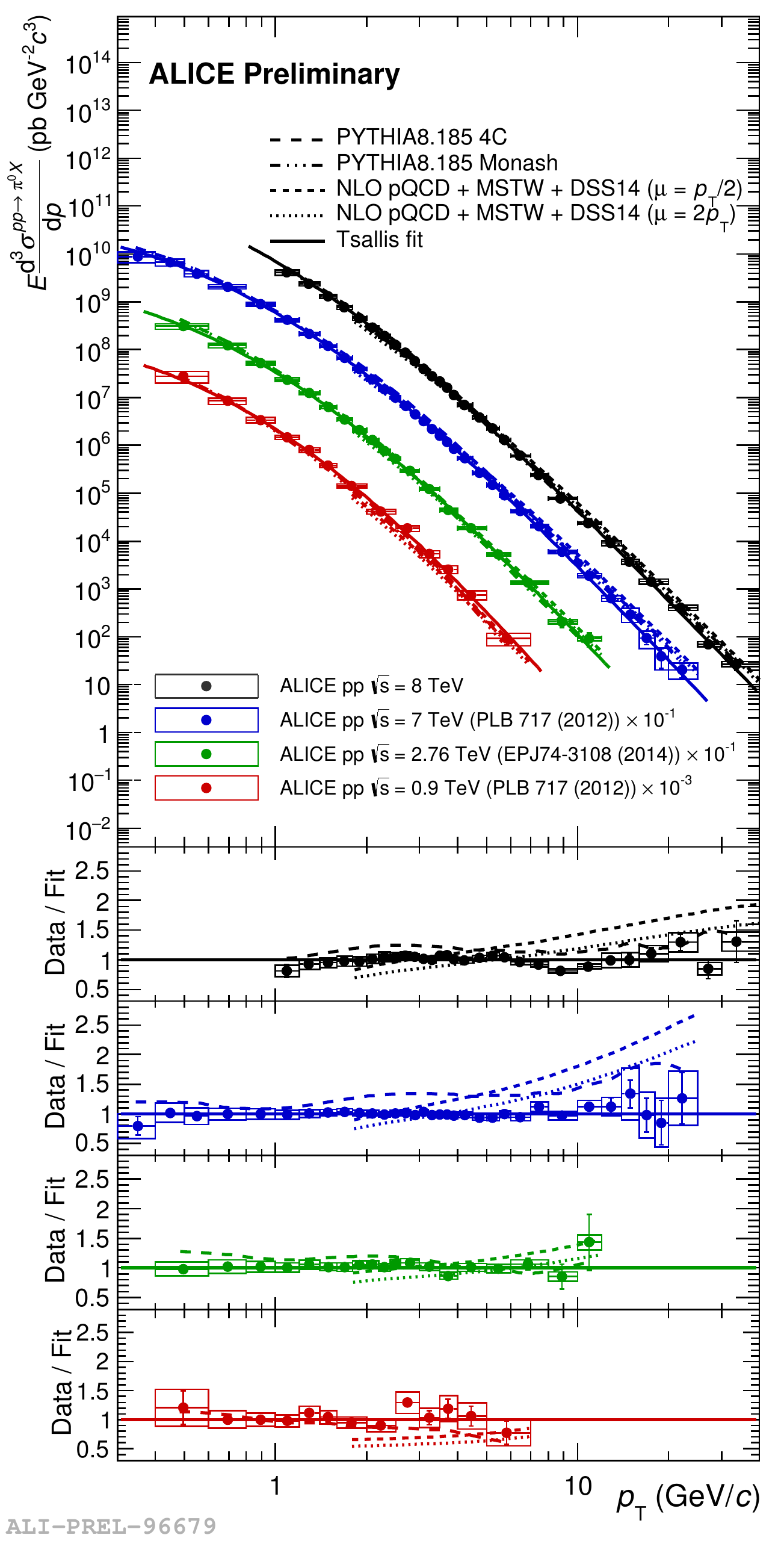}
\caption{\footnotesize $\pi^{0}$'s invariant yields measured at four $\sqrt{s}$ energies. See text for details.}\label{fig:ppyields}
 
\end{center}
% \vspace{-2cm}
\end{wrapfigure}

\vspace{-0.3cm}
\section{$\pi^{0}$ invariant yields~in~pp~collisions~at $\sqrt{s} =$ 0.9, 2.76,  7  and 8 TeV}
The ALICE experiment to this date has measured the invariant yields of $\pi^{0}$ at four ${\sqrt{s}}$. 
The measurements indicate a power law dependence at high $p_{\rm T}$~with  a measured power value of $n=6.0\pm0.1$ at $\sqrt{s}=$~2.76~TeV, a value  lower than what has been observed with measurements performed at lower ${\sqrt{s}}$ energies~\cite{Adare:2013esx}. The measurements are also compared to next-to-leading order pQCD (NLO pQCD) which use MSTW PDFs, and DSS14 fragmentation functions (FF)~\cite{deFlorian:2014xna}. These comparisons as well as the ratio are illustrated in Fig.~\ref{fig:ppyields}. 
The predictions with MSTW PDF and DSS14 FF describe the magnitude better than previous pQCD calculations~\cite{Abelev:2012cn}, however an increasing discrepancy between pQCD and the measurements is observed with increasing  ${\sqrt{s}}$ and $p_{\rm T}$.  
\vspace{-0.3cm}
\section{Invariant~yields~in~Pb-Pb~collisions compared to pp}
\begin{figure}[h]
\begin{center}
\includegraphics[scale=0.26]{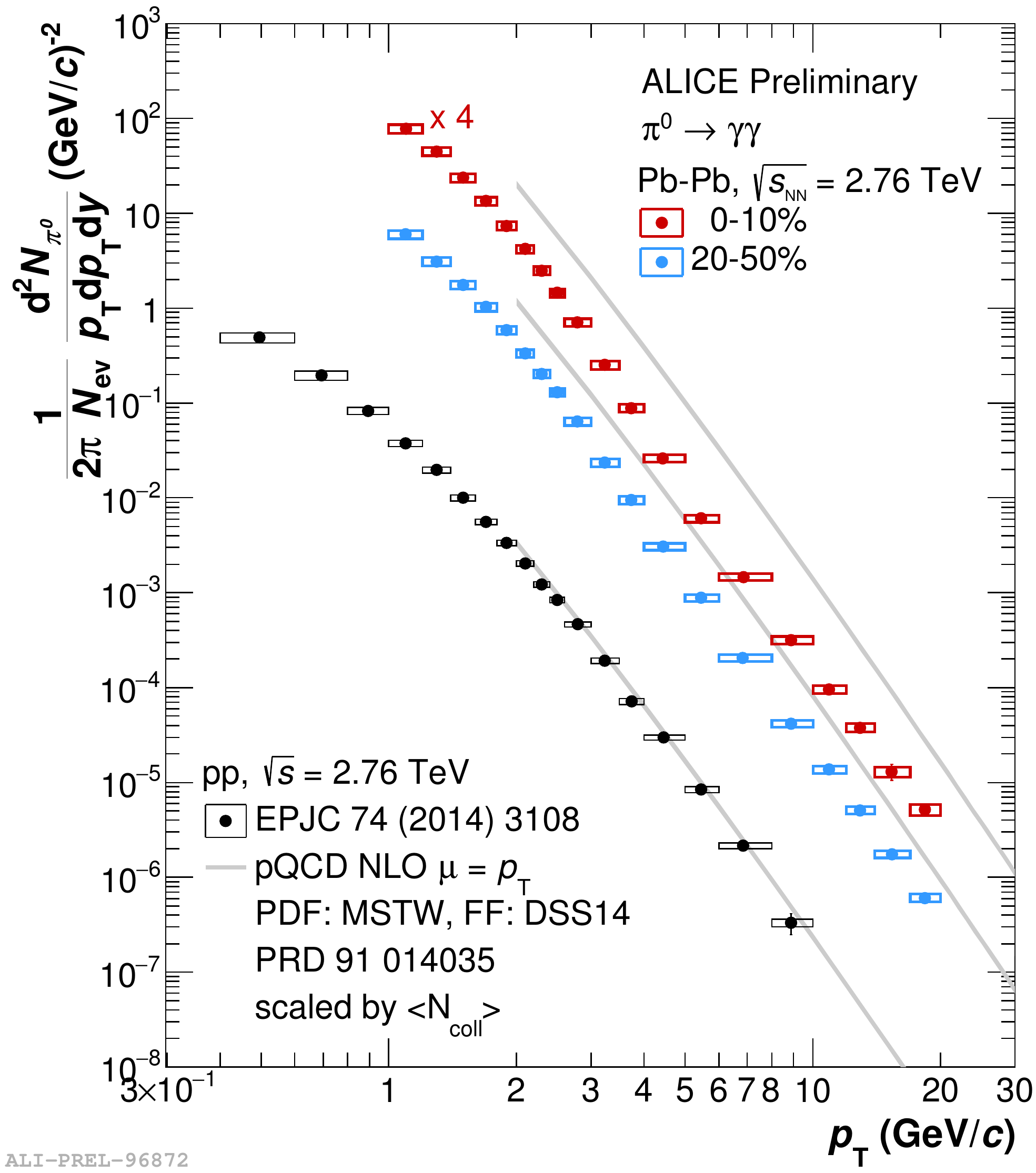}
\includegraphics[scale=0.26]{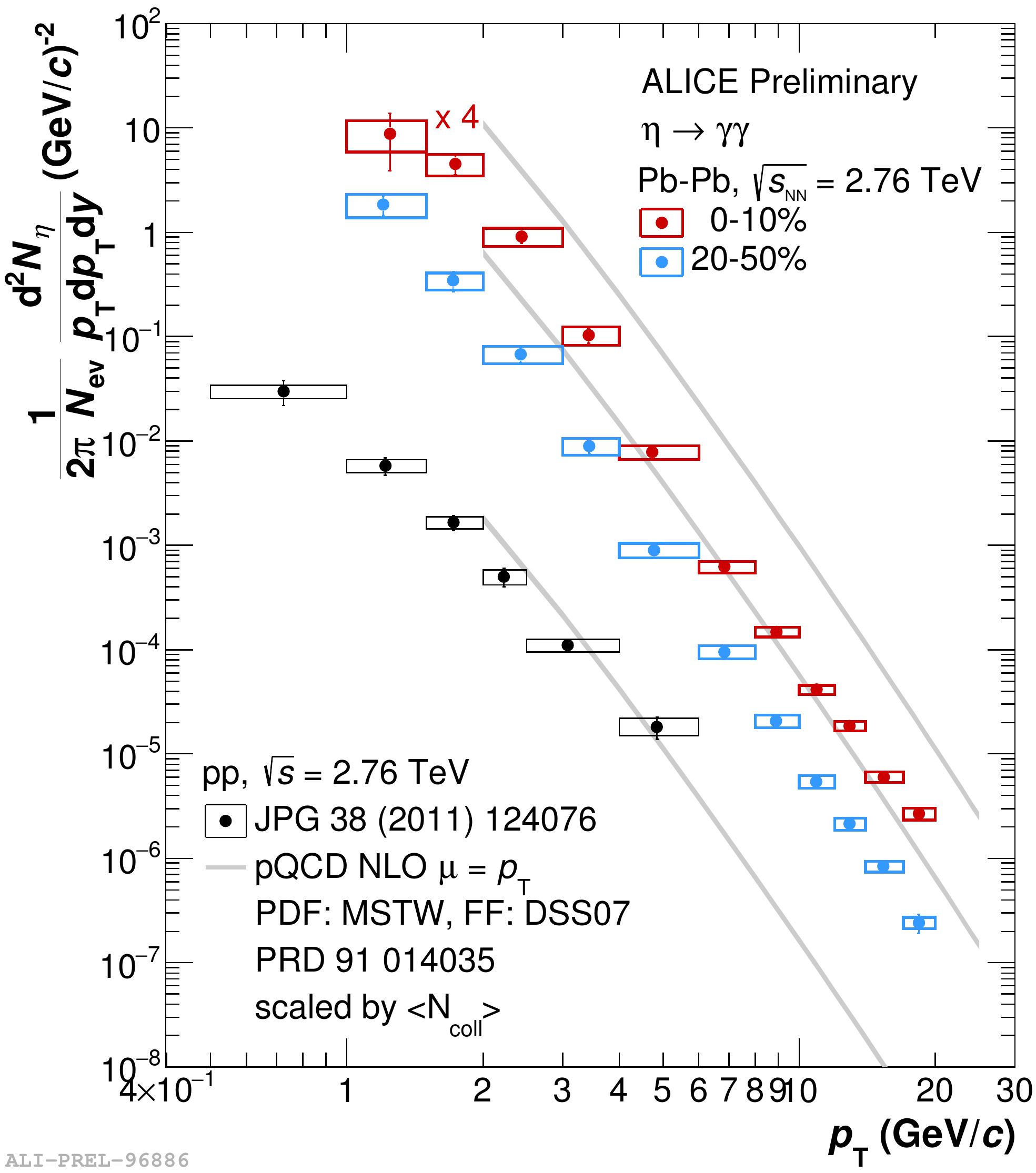}
\includegraphics[scale=0.26]{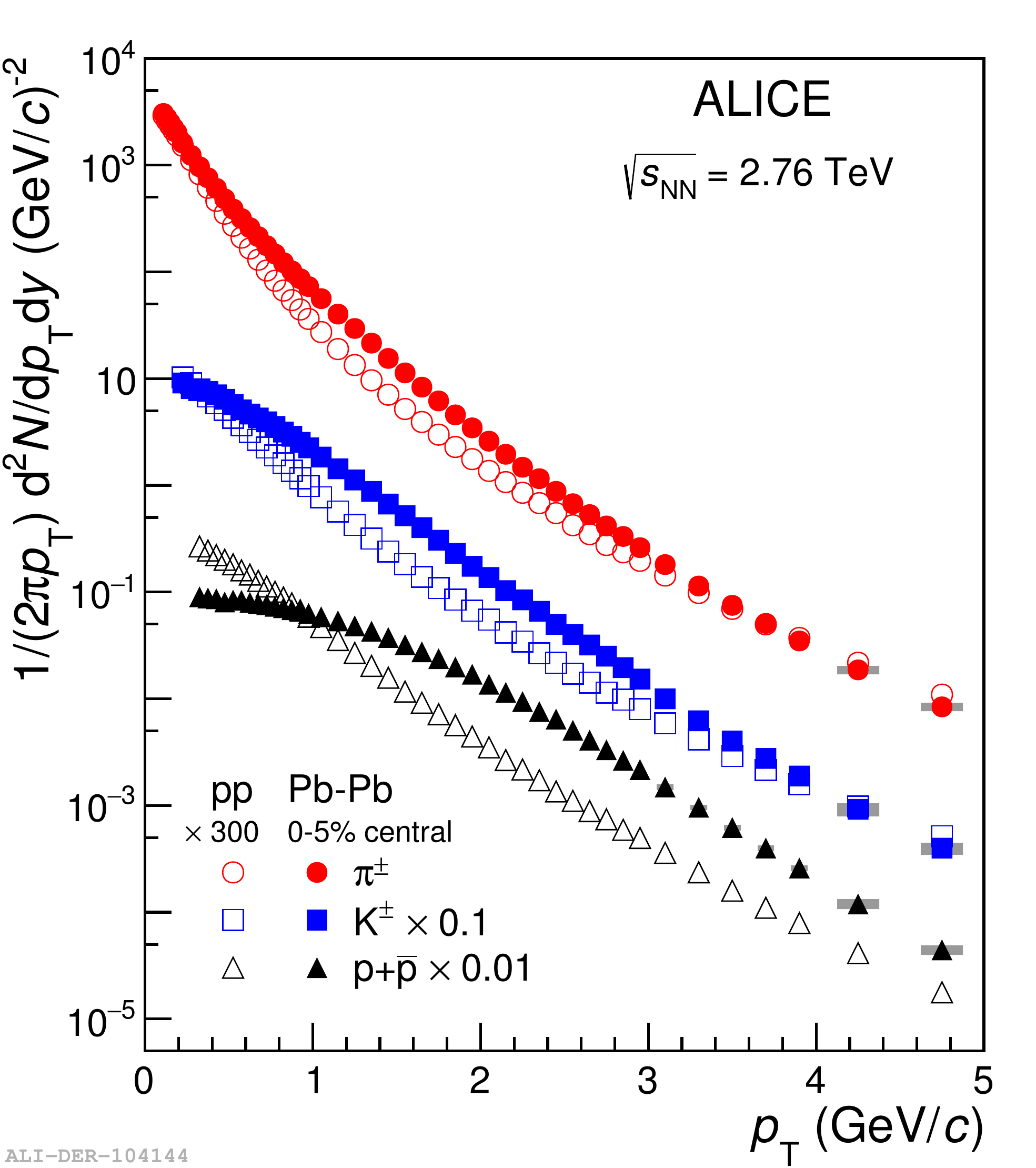}
\caption{\footnotesize Invariant yields in Pb-Pb collisions of $\pi^{0}$ (left), $\eta$ (middle) and K$^{\pm}$, $\pi^{\pm}$ (right)}\label{fig:newpbpbyields}
\end{center}
\end{figure}

Measurements of $\eta$, $\pi^{0}$, $K^{\pm}$ and $\pi^{\pm}$ mesons in HIC are presented in Fig.~\ref{fig:newpbpbyields}.
The new $\pi^{0}$ results in Fig.~\ref{fig:newpbpbyields} are from a data set offering ten times the integrated luminosity with respect to previous ALICE results\cite{Abelev:2014ypa}.
%\clearpage 
This increase of statistics allowed for an improved measurement that probes the $p_{\rm T}$~region above 12~GeV/$c$~ and up to 20~GeV/$c$. As for $\eta$ mesons (Fig.~\ref{fig:newpbpbyields}),  this the first time this particle is measured at the LHC. 
\clearpage
\vspace{0.3cm}
The $\pi^{0}$ and $\eta$ results are compared in two centrality classes to NLO pQCD predictions of the corresponding production in pp collisions~\cite{deFlorian:2014xna} scaled by $N_{coll}$. These comparisons show a suppression in the Pb-Pb data with respect to scaled NLO pQCD.  $K^{\pm}$ and $\pi^{\pm}$ in the 0-5$\%$ centrality class (Fig.~\ref{fig:newpbpbyields})  are compared to corresponding pp measurements (open markers), where a zoom to the $p_{\rm T}$ region below 5 GeV/$c$ indicate differences in the spectra with respect to pp that is attributed to radial flow~\cite{Abelev:2014laa}.

\vspace{-0.3cm}
\section{Nuclear modification factor ${R_{\rm AA}}$}
\begin{figure}[h]
\begin{center}
\vspace{0.5cm}
\includegraphics[scale=0.6]{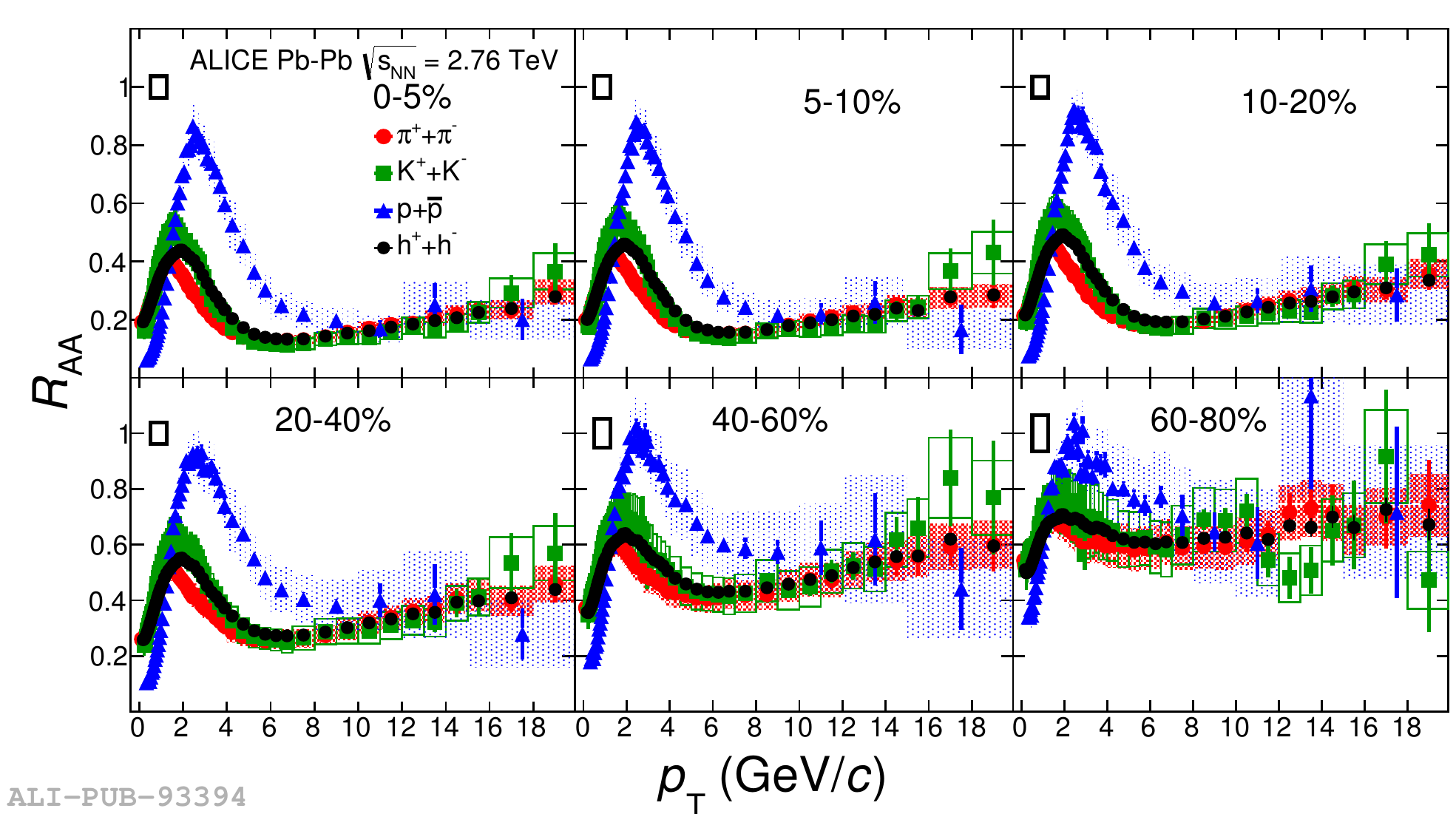}
\caption{\footnotesize  $R_{\rm AA}$ in six centrality classes. See text for details.}\label{fig:kaonsraa}
\end{center}
\end{figure}

The nuclear modification factor ${R_{\rm AA}}$ is defined as: $R_{\rm AA}(p_{\rm T})=\frac{1}{N_{coll}}\frac{dN_{\rm AA}/dp_{\rm T}}{dN_{pp}/dp_{\rm T}}$
is measured to quantify nuclear effects in A-A collisions, where production in A-A is  compared to production in scaled pp~collisions. ${N_{coll}}$ is the number of binary nucleon-nucleon collisions and is taken from Glauber Monte Carlo simulations~\cite{Abelev:2012hxa}.
Fig.~\ref{fig:kaonsraa} shows the measured $R_{\rm AA}$ for kaons and pions as well as other particles, where a large suppression in central Pb-Pb collisions is observed,  0-5$\%$: ${R_{\rm AA}}\sim$0.1 for $p_{\rm T}>$ 6 GeV/$c$, and a moderate suppression at peripheral collisions,  60-80$\%$: ${R_{\rm AA}}\sim$ 0.6 for $p_{\rm T}>$ 6 GeV/$c$. At high $p_{\rm T}$ ($>$8 GeV/$c$), all particle species exhibit a similar suppression trend which decreases torwards large $p_{\rm T}$. At low $p_{\rm T}$ the magnitude of the ${R_{\rm AA}}$ increases with the increasing particle's mass.

\vspace{-0.1cm}
\section{Particle ratios  in Pb-Pb collisions}
\begin{figure}
 \begin{center}\vspace{-0.3cm}
\includegraphics[scale=0.27]{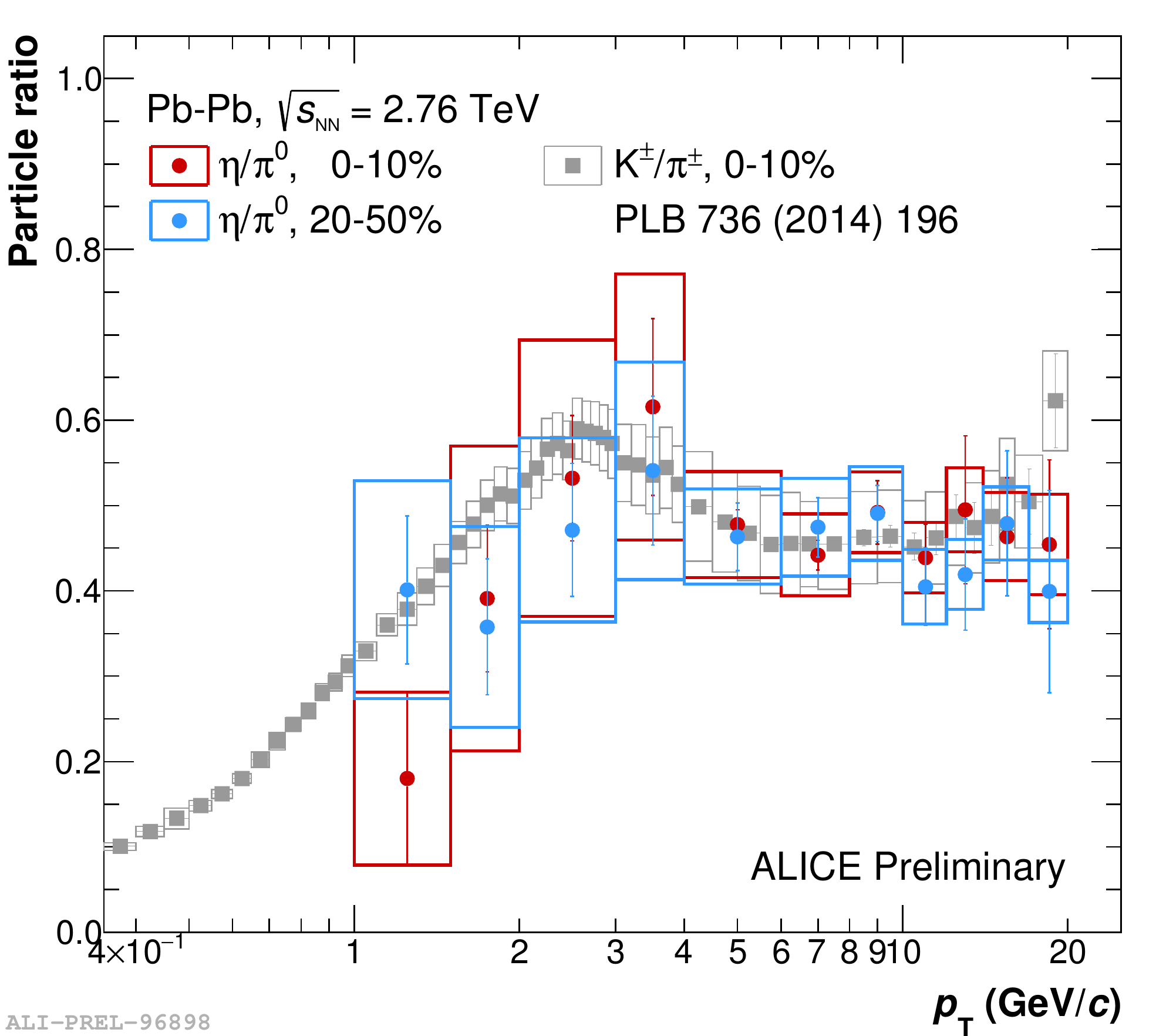}\includegraphics[scale=0.27]{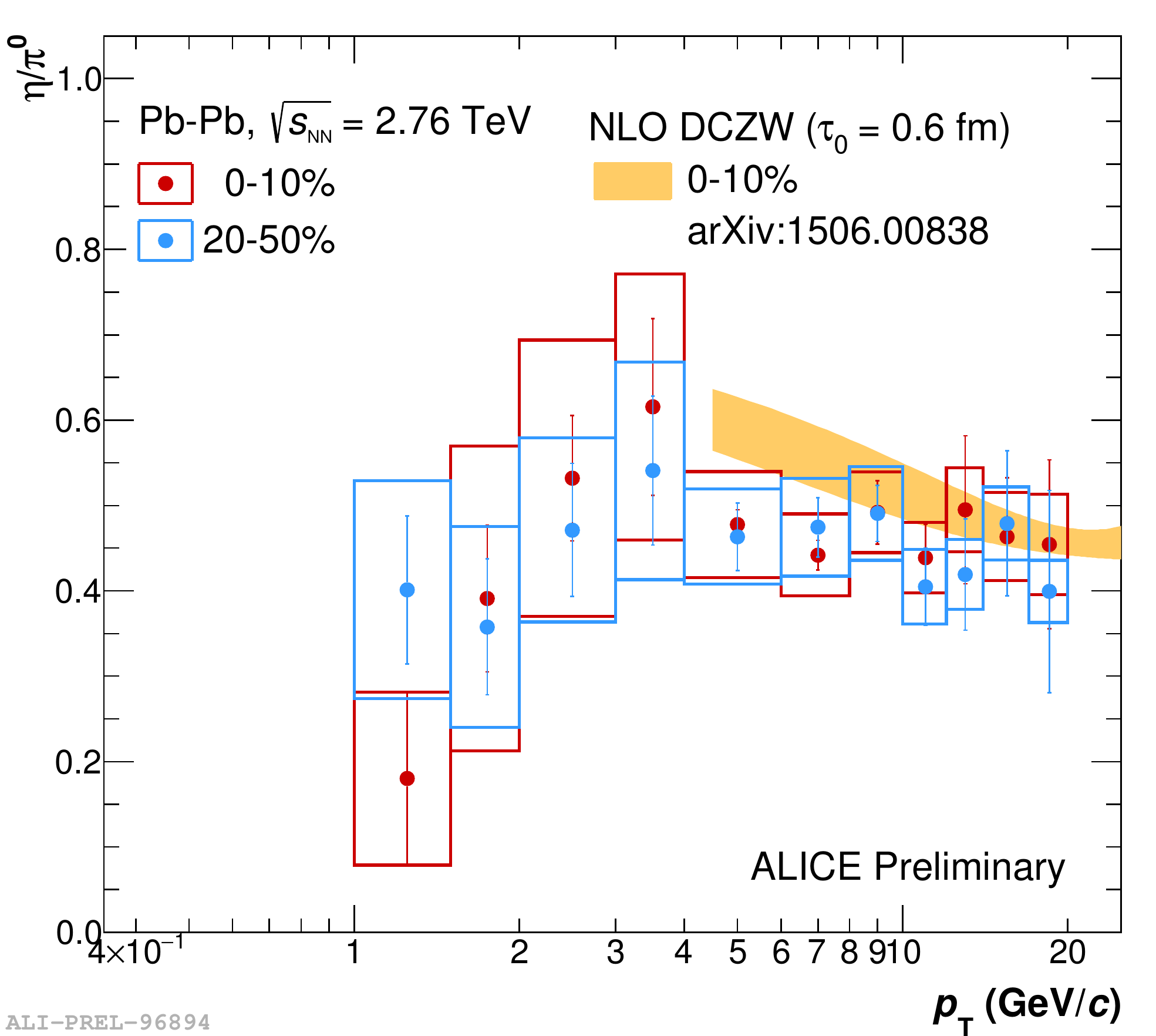}\includegraphics[scale=0.27]{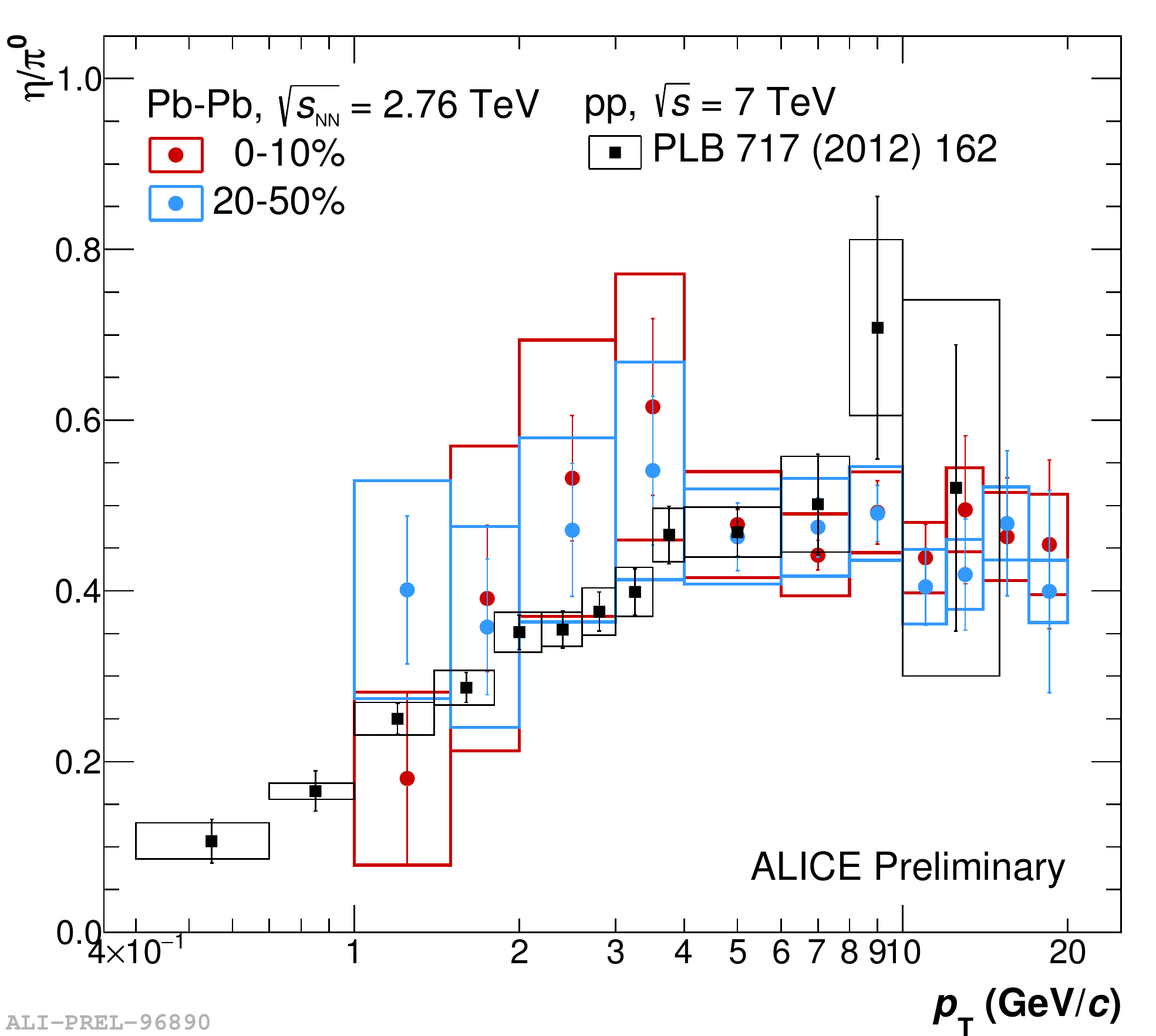}
\caption{\footnotesize $\eta/\pi^{0}$ in two centrality classes compared to $K^{\pm}/\pi^{\pm}$(left), NLO pQCD (middle) and $\eta/\pi^{0}$ measured in pp collisions (right).}\label{fig:particleratios}
\vspace{-1.0cm}
\end{center}
\end{figure}

Fig.~\ref{fig:particleratios} shows the $\eta/\pi^{0}$ in two centrality classes compared to:  $K^{\pm}/\pi^{\pm}$ at the same $\sqrt{s_{\rm NN}}$  and the $\eta/\pi^{0}$ measurement in  pp collisions at ${\sqrt{s}} =$ 7 TeV. The trend and magnitude of the $\eta/\pi^{0}$  ratio is  consistent with the $K^{\pm}/\pi^{\pm}$ ratio ~\cite{Abelev:2014laa} as well as the corresponding ratio in pp collisions at $\sqrt{s}=$ 7~TeV. Further comparisons to jet quenching NLO pQCD predictions (Fig.~\ref{fig:particleratios} middle) ~\cite{Dai:2015dxa} describe the ratio within the current uncertainties.
\vspace{-0.3cm}
\section{Summary}
ALICE measures neutral and charged mesons in a wide $p_{\rm T}$ range thanks to complementary detectors.
$\pi^{0}$, $\eta$ and $K^{\pm}$ invariant yields have been measured both in pp as well as Pb-Pb collisions by ALICE and compared to NLO pQCD.
$\pi^{0}$'s magnitude is well described by NLO pQCD calculations, however, there is a growing discrepancy as a function of $p_{\rm T}$~and with increasing ${\sqrt{s}}$ (pp results). 

A suppression of $\pi^{0}$ and $\eta$ is observed via comparisons of the Pb-Pb invariant yields to scaled pp NLO pQCD. $K^{\pm}$ and $\pi^{\pm}$ suppression is observed via  ${R_{\rm AA}}$ which is $p_{\rm T}$  and particle dependent (p$_{T}<$ 6 GeV/$c$). At high $p_{\rm T}$ ($>$ 8GeV/$c$) the value and trend of the ${R_{\rm AA}}$ is particle independent pointing that the suppression may be a partonic effect alone. At low p$_{T}$ a mass hierarchy is observed on the magnitude of the ${R_{\rm AA}}$  attributed to radial flow. 
Particle ratios measured in Pb-Pb collisions were presented and compared to jet quenching predictions as well as corresponding ratios measured in pp.  $K^{\pm}/\pi^{\pm}$~reaches a constant value for $p_{\rm T}>4$ GeV/$c$. For $p_{\rm T}$~below 4~GeV/$c$ an enhancement is observed which is centrality dependent.
 $\eta/\pi^{0}$~ reaches a constant value for $p_{\rm T}>4$ GeV/$c$. For $p_{\rm T}$~below 4~GeV/$c$ it is consistent with the pp and $K^{\pm}/\pi^{\pm}$ ratio.  At high $p_{\rm T}$ no significant differences are seen in the particle ratios between Pb-Pb and pp.

\section*{References}

\bibliography{iopart-num}

\end{document}